\documentclass[runningheads]{llncs}
\usepackage{graphicx}
\usepackage{enumitem}
\begin{document}
\title{Design of a Representation Information Repository for the Long-Term Usability of Digital Building Documents}
\titlerunning{Repository for the Long-Term Usability of Digital Building Documents}
\author{Uwe M. Borghoff \inst{1}\thanks{Corresponding author: uwe.borghoff@unibw.de}
\and Eberhard Pfeiffer \inst{2} 
\and Peter R{\"o}dig \inst{1}
%\orcidID{0000-0002-7688-2367}
%\thanks{Corresponding author.}
}
\authorrunning{U. M. Borghoff, E. Pfeiffer \& P. R{\"o}dig}
\institute{Institute for Software Technology, Department of Computer Science \\ University of the Bundeswehr Munich, Neubiberg, Germany\\
\and
Department of Civil Engineering and Environmental Sciences \\ University of the Bundeswehr Munich, Neubiberg, Germany\\
%\email{uwe.borghoff@unibw.de}\\
%\url{https://www.unibw.de/ciss} 
}
\maketitle              % typeset the header of the contribution

\begin{abstract}
The long-term usability of digital building documents is essential for the maintenance and optimization of infrastructure portfolios. It supports the preservation of building-specific knowledge and the cultural heritage hidden within. However, having to do this throughout the lifecycle of a building -- or even indefinitely -- remains a major challenge. This is especially true for organizations responsible for large collections of digital building documents, such as public administrations or archives. 
In this article,\footnote{This paper expands on some issues discussed and presented in a preliminary version at DocEng ’22 \cite{BorghoffPR22}.} 
we first describe the challenges and requirements associated with preservation tasks, and then introduce the concept of so-called representation information within BIM (Building Information Modeling). This type of information is important to give meaning to the stored bit sequences for a particular community. Then, we design a repository for representation information and introduce some so-called 23 BIMcore content elements. Finally, we focus on BIM and the construction sector and explain how the proposed repository can be used to implement the two concepts introduced in the ISO reference model OAIS (Open Archival Information System), namely the representation information and the context information, as well as the concept of significant properties, which has not yet been explicitly modeled in OAIS.

\end{abstract}

\section{Introduction and Challenges}
Resource scarcity, cost increases, climate change, and demographic change pose major challenges for the public and private construction sector. However, digitization of these sectors can help to manage the associated tasks, especially if the entire life cycle of buildings, including demolition and recycling of materials, is considered on the basis of digital information, or digital twins in the future, see, for instance, \cite{madubuike2022review} or
\cite{alshammari2021cybersecurity}.
In particular, the permanent usability of digital building information can offer concrete benefits: 
\begin{itemize}
\item
the avoidance of the loss or degradation of information, and the repeated and cost-intensive creation of new data in the case of renovations, repurposing, repairs, deconstruction planning or waste management; 
\item
the preservation and expansion of knowledge for use in new projects;
\item 
the economic realization of complete long-term analyses of buildings or entire portfolios; 
\item
the improvement of computational models or their parameters on the basis of data from long-term monitoring; 
\item
planning and simulation of safety and emergency measures every time in the life cycle; 
\item
the preservation of cultural built heritage through virtual structures, especially when physical preservation or restoration is not possible;
\item 
the fulfillment of all formal requirements for documentation and archiving. 
\end{itemize}

Ensuring the long-term usability of digital information and safeguarding the digital cultural heritage is one of the grand challenges of information technology. The peculiarities and boundary conditions of the construction sector and public administration increase the challenges significantly. Numerous problem areas that relate specifically to the objective of this paper can be identified, i.e. 
\begin{itemize}
\item
the rapid obsolescence of formats due to short innovation cycles for hardware, software, and human-machine interaction; 
\item
the large number of formats for data, files, file systems, database schemas, protocols for data exchange as well as the insufficient documentation of their context; 
\item
incomplete standardization, proprietary formats, and imprecise specifications; 
\item
an increasing sharing of work and distribution of related storage and processing systems.
\end{itemize}

\subsection{Building Information Modeling (BIM) and the Digital Transformation}
Electronic data processing has a long tradition in engineering and it was a long way to the sophisticated calculation methods available today. After around 1950, when the first freely programmable computing systems developed by the civil engineer Konrad Zuse became available, it was initially a question of finding out how traditional manual calculation methods tailored to slide rules could be replaced by computer-adapted procedures. The increasing availability of computer capacity at universities and large-scale research facilities led to the rapid development of new numerical methods worldwide, with which complex structures and systems of equations with several tens of thousands of unknowns could be calculated reliably and quickly despite limited computer resources.

\begin{figure}[ht]
\includegraphics[width=\textwidth]{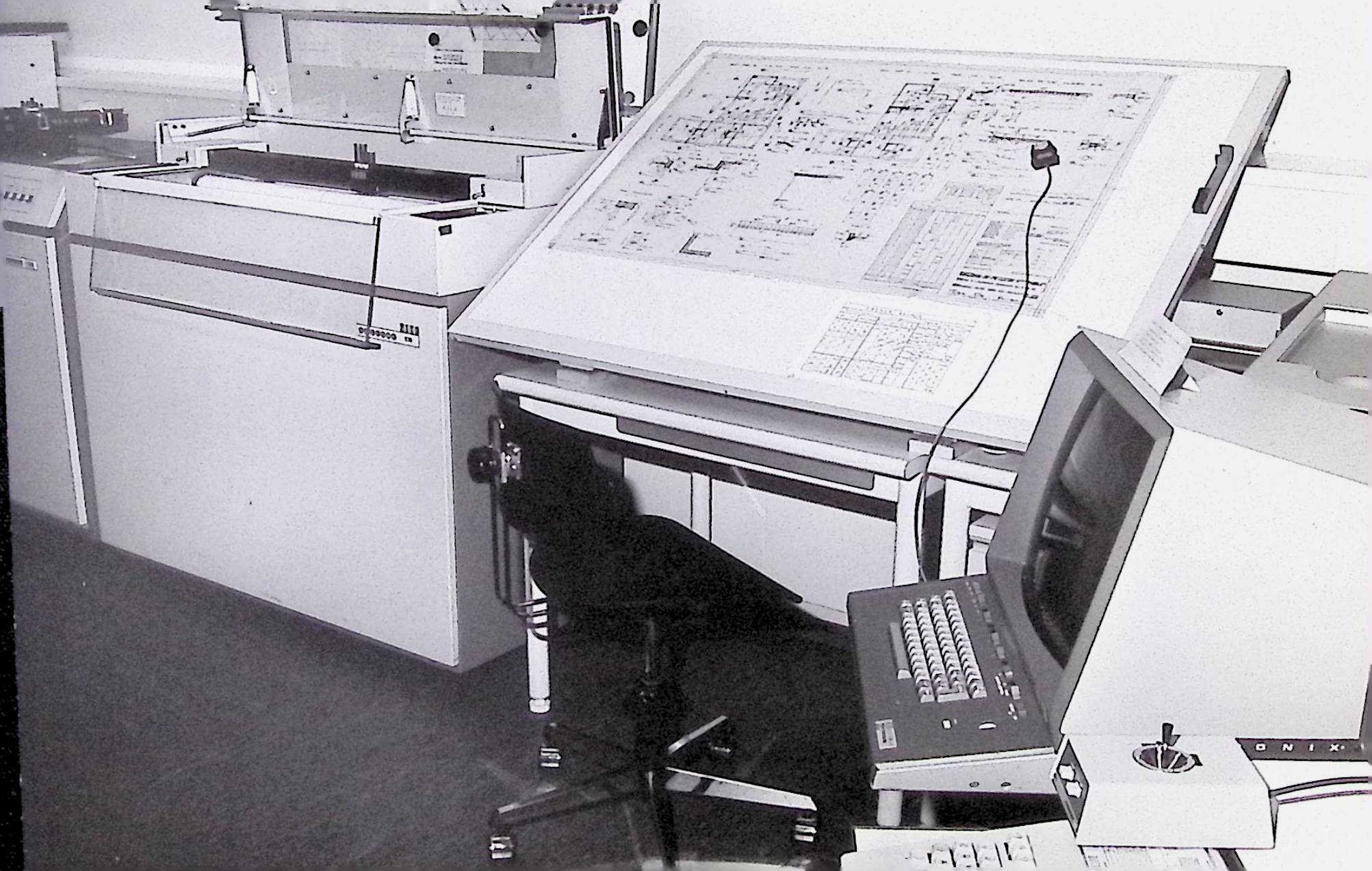}
\caption{Civil engineer workplace around 1978.} \label{Fig1}
\end{figure}

Human interaction with computers was long time limited by punched tapes, punched cards, strictly formatted and therefore error prone input formats for commands and data, and often long printouts that had to be laboriously evaluated. Terminals, later with graphical functions and thumb wheels or joystick for cursor operations, digitizer tablets and plotters gradually made work more efficient. FIG.~\ref{Fig1} shows such a workplace at our IT-lab around 1978, consisting of a tablet for digitizing technical drawings and a menu field for the specification of datasets and for program control with symbolic buttons. The configuration also included a vector-based display with hardcopy unit, a pen plotter with a tape unit for recording graphic output or possibly input of saved drawings, and still a punch card-reader and a high-speed printer. The computer was a SIEMENS Process 330 system with an alphanumeric console and units for magnetic tapes and removable disks. A FORTRAN compiler and the Tektronix PLOT-10 graphical software served as the development platform for our advanced software library for structural and fluid dynamic analysis. All program developments from this time and to a large extent also from earlier software projects were continuously documented with flow charts, memory plans and source codes. This made it much easier and faster to later rewrite it in other programming languages and to port it to almost all computer platforms, from mainframes to embedded or mobile systems. This is a best practice example of how changing development teams over time could benefit from well-documented prior work when reusing code.

Back then, construction documents were usually paper-based, with computer-generated drawings and extensive lists attached, but in some cases digital media such as floppy disks with input and output files were also included.

BIM has emerged as the single and preferred method for planning and managing construction projects based on a virtual representation of a building with all the features required for integrating the specific models of all the disciplines involved. However, the introduction of BIM brings additional challenges for long-term data and document management. Advanced analysis methods based on digital building models and innovative monitoring and production models (e.g.\ virtual spare parts based on 3D printing formats) lead to highly specialized formats and, in some cases, large volumes of data. In the context of preservation (e.g., heritage preservation), extended data structures for buildings and their semantic enrichment (linked data, discipline-specific ontologies) must be taken into account. In addition, the management and adaptation of complex and meaningful digital models require the introduction of metamodels, especially when it comes to schema extensions. Another critical case is the derivation of virtual models or documents based on complex numerical algorithms. It is clear that the digital transformation is changing the traditional concept of documents and will lead to new challenges for long-term archiving due to increasing virtualization, continuous updating over long periods of time and the embedding of semantically enriched complex 3D models, as is already partly the case in PDF/E-1.\footnote{https://www.iso.org/standard/42274.html}

\subsection{Gaps -- what is missing?}
We have found sophisticated solutions for the long-term usability of digital documents for technical and scientific areas, but these do not take the construction sector into account, or only to a limited extent. The implemented repositories or registers lack depth of representation and contextual information or construction-specific content; in particular, there is a lack of identification and description of significant properties, and the current approaches do not take into account the new challenges posed by BIM and digital transformation described above. In addition, some repository projects were canceled or stopped after the prototype phase. The lack of commitments or guarantees regarding the durability of the services poses risks that are difficult to assess.

\section{Related Work}
For about a quarter of a century, archives, libraries, scientific research institutions with large and valuable data collections, but also the industry as well as standardization organizations have been dealing with the challenges of digital long-term archiving. The broadly discussed and accepted ISO reference model OAIS (Open archival information system) \cite{book2002consultative} is a milestone for the preservation community.

\begin{figure}[ht]
\includegraphics[width=\textwidth]{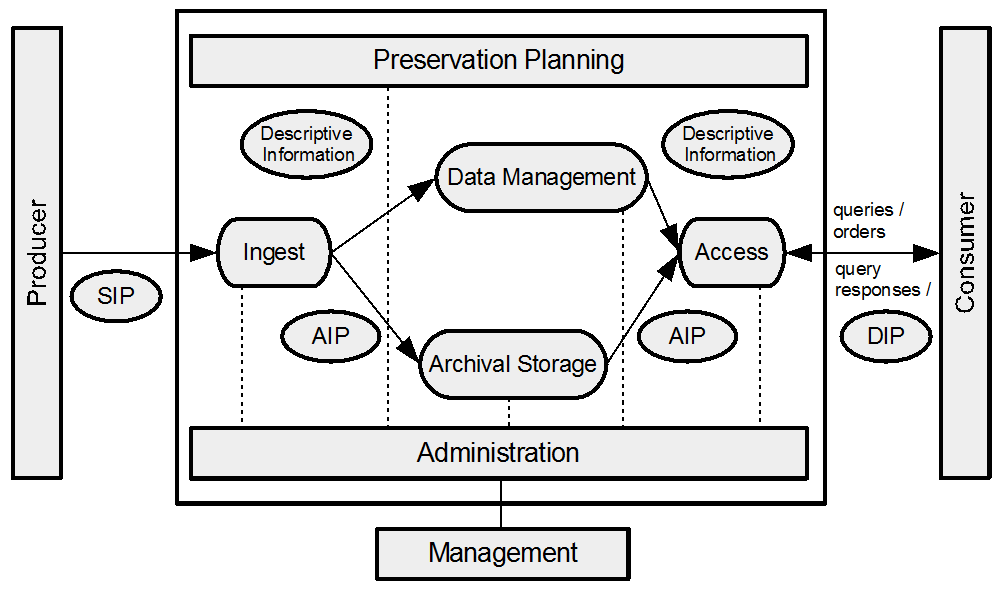}
\caption{Functional and data entities of the ISO reference model OAIS (Open Archival Information System).} \label{Fig2}
\end{figure}

OAIS describes the essential components of an archiving system for long-term information preservation at the conceptual level. It considers data as information packages that contain metadata (e.g., about the data format) in addition to the content. The packages submitted by the producer for archiving are called {\em submission information packages} (SIP). They are converted into {\em archive information packages} (AIP) upon submission. These contain a variety of information, including the history of previous processing steps as well as necessary migration data and other attributes required for integrity and authenticity. Searching and access by users also takes place via dedicated information packages, the so-called {\em dissemination information packages} (DIP); see also FIG.~\ref{Fig2}.

Key terms include so-called {\em representation information} (RI), \cite{farghaly2022construction} which includes all the information needed to interpret a bit sequence in order to convey a meaningful perception to a particular community; see also \cite{BohneRB11}. Format descriptions are an important form of representation information and therefore, format registries play an important role in the community. A further concept of OAIS is context information, which contributes to a deeper understanding of digital content; e.g., user manuals or specifications of measuring devices.

Taking into account the provision of relevant content or a conceptual and technical reusability, we have considered the following format registers: 

(i)
GDFR/UDFR (Global Digital Format Registry/Unified Digital Format Registry),\footnote{https://preservation.library.harvard.edu/digital-preservation/projects-and-initiatives} 

(ii)
PRONOM of the National Archive UK,\footnote{https://www.nationalarchives.gov.uk/PRONOM/Default.aspx} and 

(iii)
the LOC (Library of Congress, USA) offerings called Sustainability of Digital Formats - Planning for Library of Congress Collections, which also contain evaluation criteria for the sustainability of formats,\footnote{https://www.loc.gov/preservation/digital/formats} 

(iv)
KAD (Catalog of Archival File Formats) with evaluation systematics for archives from the Swiss institution KOST (Koordinationsstelle für die dauerhafte Archivierung elektronischer Unterlagen),\footnote{https://kost-ceco.ch/cms/kad\_main\_de.html} and 

(v)
the File Format Database from Sharpened Productions, USA.\footnote{https://fileinfo.com}

Two developments considered the concept of representation information explicitly: 

(vi) The RRoRIfE (Registry/Repository of Representation Information for Engineering),\footnote{http://rorife.sourceforge.net} which was intended as a tool for preservation planning in CAD, CAM, and CAE, and 

(vii) the DCC (Digital Curation Centre) Representation Information Repository, which was intended for a broader scope.\footnote{https://www.dcc.ac.uk/guidance/briefing-papers/technology-watch-papers/caspar}

The best-known tool in the field of memory organizations for identifying and validating different file formats is JHOVE (JSTOR/Harvard Object Validation Environment).\footnote{https://jhove.openpreservation.org/documentation}

Another means with a long tradition for recognizing a variety of file or file system formats is the UNIX program file.\footnote{https://pubs.opengroup.org/onlinepubs/9699919799/utilities/file.html}
DPC (Digital Preservation Coalition, UK) has been dealing with preserving CAD again and issued a short guide in 2021, which contains concise descriptions of a selection of formats commonly used in the construction sector.\footnote{http://doi.org/10.7207/twgn21-15}

\section{First Results}
Borghoff et al.\ \cite{BorghoffPR22} present a holistic and standard-based concept for the long-term usability of datasets and documents common to building authorities and other stakeholders that are responsible for large digital assets. 

\begin{figure}[ht]
\includegraphics[width=\textwidth]{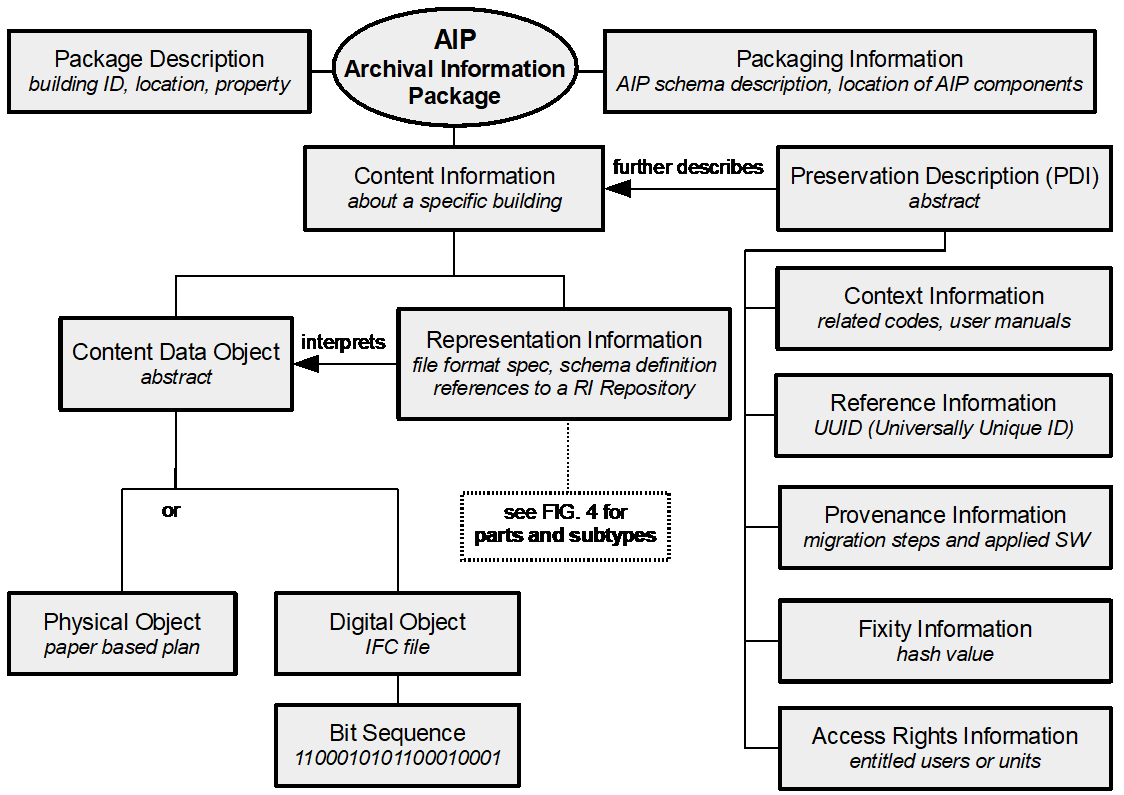}
\caption{OAIS's data entity Archival Information Package (AIP) and its components.} \label{Fig3}
\end{figure}

Here, we refine an important conceptual element of OAIS, namely the Archival Information Package (AIP). 
In particular, we focus on the basic concept of data objects (bit sequences) and their interpretation by {\em representation information}, see also FIG.~\ref{Fig3}. We describe the concept of representation information and present a design for the development of a repository for this type of information, in particular for the storage of BIM documents.

\begin{figure}[ht]
\includegraphics[width=\textwidth]{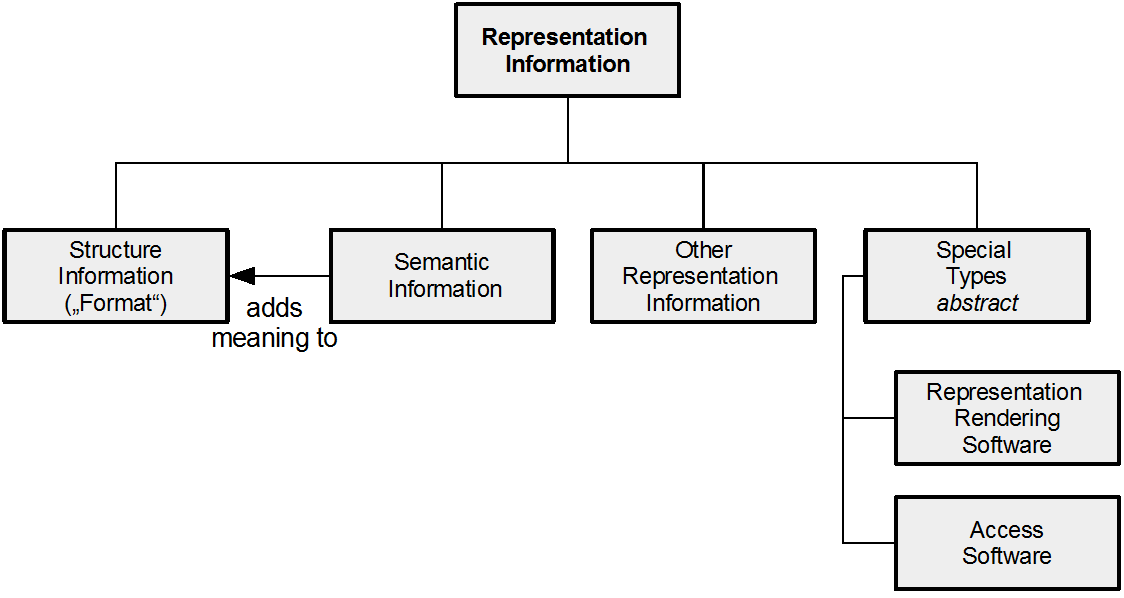}
\caption{Parts and subtypes of AIP's component Representation Information (RI).} \label{Fig4}
\end{figure}

{\em Representation information} (RI) as a key concept of OAIS is responsible for giving the stored bit sequences meaning for the designated communities. As FIG.~\ref{Fig4} depicts, the standard distinguishes between three subtypes of RI: (1) {\em structure information} should describe the elementary structure of the bit sequences so that all subsequences can be identified and assigned to primitive data types and data structures and further on to higher level concepts; e.g., four subsequences with the length of four bytes in endian-order each may represent four numbers, which in turn represent the coordinates of the starting and ending point of a 2D-line. (2) {\em semantic information} should provide additional meaning to all the elements described by the structural information, e.g., a set of lines in a defined order and coordinate system may represent a floor plan of a building at a defined scale. (3) the so-called {\em other representation information}, which should complete the {\em representation information} in order to understand a data object as comprehensively as possible. OAIS lists these examples: software, algorithms, encryption, and written instructions. In the case of our floor plan, an algorithm could incorporate knowledge about what should happen when one line is moved. In addition, OAIS defines two special types of RI: 
\begin{itemize}
\item
{\em representation rendering software} must be able to display digitally coded RI in an understandable form. 
\item
{\em access software} presents some or all of the information represented by a digital object to humans or systems. 
\end{itemize}

Additionally, the digital preservation community has introduced and broadly discussed the concept of {\em significant properties}, which is not explicitly described in OAIS, but we think that the following definition is suitable in our context: The characteristics of an Information Object that must be maintained over time to ensure its continued access, use, and meaning, and its capacity to be accepted as evidence of what it purports to record. For details, refer to \cite{van2018significant}.

\section{The Repository}
Our research and discussions to date have led to the identification of use cases and, building on this, to an initial definition of conceptual content elements for an RI repository. Examples in the specific BIM context are used to explain the concept and illustrate how such a repository could be applied.

\subsection{Use Cases}
We have found that it is desirable to address the following five major roles and use cases: 
\begin{itemize}
\item
{\em data and document producer} (SIPs as output) concerned with long-term preservation for legal, economic or cultural purposes; 
\item
{\em construction experts as data and document consumer} (DIPs as input) for various levels of reuse where some cases require a high degree of precision and reliability, e.g., nuclear decommissioning or the restoration of culturally valuable buildings; 
\item
the {\em management and administration of archives} for assessing the quality of formats, semantic and other representation information in order to support preservation planning as well as daily tasks like the construction of AIPs; in this use case, the focus is on the permanent preserving of information independently of a construction's lifecycle;
\item
{\em computer experts} who want to evaluate the design, reuse existing code or implement preservation tools such as emulators or converters for format migrations; 
\item
{\em computer and other technical oriented historians}, as electronic data processing has a long tradition, especially in civil and mechanical engineering.
\end{itemize}

\subsection{The BIMcore Content}
To cover the listed use cases, we propose the following 23 conceptual BIMcore content elements,\footnote{In analogy to the Dublin core, which contains metadata elements for describing Internet documents (see https://www.dublincore.org/), we call our 23 content elements the BIMcore. It contains a set of simple metadata elements for describing BIM data.} divided into three categories.

\subsubsection{Structural and semantic features.}
We combine structural and semantic aspects, as file formats can also contain semantic information in the sense of OAIS. Some formats used in the building sector can even be compared to semantic databases, as they explicitly represent domain-specific entities and relationships. Examples are IFC or multi-model/information container formats, as in ISO 21597-1:2020(en),\footnote{https://www.iso.org/standard/74389.html / Information container for linked document delivery -- Exchange specification -- Part 1: Container} which are intended to establish semantic interoperability between data with different formats and domain affiliations. Therefore, we need as content elements: 

\begin{enumerate}
\item name, identifier and classifications of a format; 
\item version of a format or reference to other versions;
\item references to already existing repositories or registries;
\item description of forward and backward compatibility issues; 
\item facts concerning the openness of a format, esp., the availability of specifications; 
\item description of rights and intellectual property to the formats and, if available, tools that allow reengineering; 
\item description of syntax (formal structure) and semantics (meaning of the individual elements and their relationships) of a format; 
\item encoding of fundamental format elements such as numbers, texts, colors; 
\item description of compression, data reduction and encryption of elements or whole files; 
\item references to external standards or documents used to specify the format or parts of it, e.g., from the following sources: ECMA, IEEE, ISO, ITU, JPEG, W3C, ICC (International Color Consortium); 
\item self-documentation capabilities of a format, e. g., the ability to include metadata; 
\item resolvability of external references (materialization); 
\item description of dependencies of a format on software and hardware to access or process the content. This is especially the case when devices, e.g., special printers, are controlled;
\item description of options for tailoring a format to specific purposes or contexts;
\item description of background and fundamentals; associated descriptions or sources may provide {\em other representation information} as defined by OAIS.
\end{enumerate}

\subsubsection{Tools for file format handling.}
Such tools should support the interpretation and quality assurance of Content Information as well as the construction and long-term maintenance of AIPs. It is clear that the processes as defined in OAIS need to be automated due to the large size of the digital collections. Additionally, we introduce the following content elements:

\begin{enumerate}[resume]
\item description of software tools for the recognition of formats; 
\item description of software tools for the verification of formats and, if applicable, tailored versions against their specification; 
\item description of software tools for inspection and (simplified) presentation of the content (see access software as a special type of RI in FIG.~\ref{Fig4}).
\end{enumerate}

\subsubsection{Contexts.}
This category should provide information that is necessary for a deeper and a long-term understanding of the digital content and for preservation planning, e.g., for assessing file format migrations, in particular, with regard to {\em significant properties}. 

Although {\em context information} is a separate conceptual element in OAIS/AIP (see FIG.~\ref{Fig3}), we propose to integrate such information into a RI repository because of the complexity of content information and the problems in distinguishing between {\em representation information} and {\em context information}. Additionally, we suggest to assign the technical aspects of {\em provenance information} to {\em context information}, for example, specification documents of instruments like sensors as an origin or source of generated data. Indeed, OAIS states that provenance could be viewed as a special type of {\em context information}.

However, we have to consider that parts of the context information may depend on organizational, local or project-specific circumstances, which in addition requires individualized repositories or an explicit integration into AIPs. For example, project manuals are a good source for describing the context of a construction project since they typically contain applied standards and guidelines as well as specifications for reporting and project documentation. The developers of the format IFC have also recognized the importance of context information and therefore introduced an entity called IfcContext which is the generalization of a project context in which objects, type objects, property sets, and properties are defined. IFC is an example for a format that can incorporate several components of an AIP in a standardized way. To make the list complete, we propose five more content elements:

\begin{enumerate}[resume]
\item description of typical contexts of origin and significant properties of typical application scenarios to be preserved; 
\item incorporation of building-specific {\em context information};
\item references to repositories containing use cases or even significant properties; for example, van Veenendaal et al.\ \cite{van2018significant} consider their work as a starting point for a shared, community-owned database with explicit knowledge on significant properties. Otherwise, use cases can help to identify and describe significant properties. So, UCMS (Use Case Management Service) offered by buildingSMART International enables the capture, specification and exchange of best practices and makes them accessible to the entire built asset industry;\footnote{https://ucm.buildingsmart.org/}
\item description of acceptance and frequency of use of a format within a specific context;
\item description or sources of background and fundamentals.
\end{enumerate}

Finally, it should be noted that the conceptual content elements of the repository will largely consist of digital objects and therefore also require RI for their long-term understanding. {\em Representation rendering software} as a special type of RI should be able to present RI in an understandable way in the long-term. Of course, {\em access software} as another special type of RI needs additional considerations.

\subsection{Application of the RI Repository in a BIM Context}
In this section, we explain how the proposed RI repository can be used to implement the two OAIS concepts of {\em representation information} and {\em context information}, as well as the concept of significant properties, which is not explicitly modeled in OAIS. We focus here on the specifics of the construction sector and BIM; therefore, we omit details found in general purpose file format repositories, such as IDs, rights and intellectual property issues, or metadata for managing such a repository.

Even medium-sized construction projects require a large number of digital documents (five hundred or more) reflecting the information and documentation needs of the parties involved, such as owners, architects, engineers, contractors, project controllers, and public administrators. Each of them uses their own tools with specific formats, schemas and configurations. We have found that common and widely used tools and formats are in use, especially for word processing and spreadsheets, but in practice, domain-specific and sometimes very complex formats from different vendors dominate. Since buildings generally have a long life span, digital assets also contain obsolete formats. The associated heterogeneity leads to problems with data exchange, data consistency and correct interpretation, especially in the long term. BIM is a widely accepted approach to minimizing these shortcomings, with the semantically enriched and standardized {\em industry foundation classes} (IFC) format, including evolving co-standards and regulations, playing a key role; see Farghaly et al.\ \cite{farghaly2022construction} for a recent application. However, digital collections, especially in public administrations, still contain digital objects with proprietary, outdated or semantically deficient formats, such as those used for so-called plot files or TIFF files, which are retro-digitized analog plans.

The following example illustrates how the proposed layout of a repository for {\em representation information} can be applied. We start with an IFC file that is to be kept for the operation of a building and for reuse. Since IFC is now a widely used format, information about it can be found in PRONOM\footnote{https://www.nationalarchives.gov.uk/aboutapps/pronom/tools.htm} or on the LOC web pages.\footnote{https://www.loc.gov/preservation/digital/formats/intro/intro.shtml}

Because IFC covers various disciplines and aspects at multiple levels of detail and development, it is common to tailor the format to specific purposes by extending or limiting the standard schema, e.g., to meet information needs for verifying and documenting compliance with fire codes.

Now contextual information comes into play as another concept of OAIS in our use case. To fully understand the content and logic represented by such a customized format, we need to know exactly what the underlying rules are and how their content can be reliably accessed -- even decades later. Therefore, we also need {\em representation information} (OAIS: {\em representation rendering software} as a special kind of {\em representation information}) when the referenced context documents are digital objects. Fortunately, these types of documents are generally encoded in a relatively simple, common, and standardized format. It would be good practice to keep such general-purpose documents that can provide {\em context information} in a long-term managed repository. However, {\em context information} or {\em representation information}, including written instructions or algorithms (OAIS: {\em other representation information} as a subtype of {\em representation information}) specific to a project or document must be integrated into an AIP, the structure of which is shown in FIG.~\ref{Fig3}. Since IFC is a rather complex format, it is desirable to have formalized and standardized tools available for customization. Such a tool, called IDS (Information Delivery Service), is under development, and the underlying file format is another candidate for an RI repository. An IDS editor is an example of {\em access software} as another specific type of {\em representation information} of the OAIS (Content Element 12). A typical use case for IDS is, for example, the adaptation of IFC building models to the requirements of fire protection (Content Element 18). In this use case, the adaptation rules reflect the {\em significant properties} of such an information object, since these rules describe fairly directly the characteristics that must be maintained over time (Content Element 18). However, identifying and defining significant properties can be quite a difficult task, especially when complex architectural or technical drawings and models for structural analysis, fluid mechanics or similar topics are to be stored for later reuse.

\section{Conclusion}
This paper illustrates the benefits and challenges of long-term usability of digital documents and datasets in a complex application domain, and then presents an approach to a key problem: preserving the meaning of digital objects with the support of a repository that provides {\em representation information} and {\em context information}. We recognize that the design, data collection, and long-term operation and maintenance of such a repository is a challenging task that can only be accomplished in a cooperative manner under the direction and responsibility of a memory organization or network of memory organizations. 

During our research, we found that a better theoretical foundation would be helpful, especially with respect to first, RI including their subtypes, and second, the inclusion of {\em significant properties} and the two OAIS conceptual elements of context and provenance information to give a repository a clear design. Another arduous task is to find and use all the knowledge sources scattered all over the world to populate such a repository with useful information. We present a metadata set, which we call the BIMcore, that will be helpful for the necessary fine structuring during the capture process. Finally, we are convinced that our approach could be useful for other scientific and technical disciplines as well.

\section*{Acknowledgement}
Our work on building authorities is in part funded by the Federal Ministry of the Interior, Building and Community (BMI) Grant Ref: SWD 10.08.18.7-15.36 under the Future Building programme. 
We thank our partners from Vintage Computing Lab (VCL)\footnote{https://www.vclab.de} and Cray-Cyber.org.\footnote{https://cray-cyber.org}
They support the operation of our datArena as an institution of the University of the Bundeswehr Munich, which is dedicated to the preservation of the digital cultural heritage. The datArena has an extraordinary pool of hardware, software and documentation.\footnote{https://www.unibw.de/inf2/forschung/forschungsthemen/datarena}


%

\end{document}